\documentclass[sigconf]{acmart}

\AtBeginDocument{%
  \providecommand\BibTeX{{%
    \normalfont B\kern-0.5em{\scshape i\kern-0.25em b}\kern-0.8em\TeX}}}

\copyrightyear{2020} 
\acmYear{2020} 
\setcopyright{acmcopyright}
\acmConference[KDD '20] {26th ACM SIGKDD Conference on Knowledge Discovery and Data Mining}{August 23--27, 2020}{Virtual Event, USA}
\acmBooktitle{26th ACM SIGKDD Conference on Knowledge Discovery and Data Mining (KDD '20), August 23--27, 2020, Virtual Event, USA}
\acmPrice{15.00}
\acmISBN{978-1-4503-7998-4/20/08} 
\usepackage{algorithmic}
\newcommand{\eat}[1]{}

\usepackage{setspace,multirow}
\usepackage{algorithm}
\begin{document}

\fancyhead{}
\title{A knowledge transfer model for COVID-19 predicting and non-pharmaceutical intervention simulation}


\author{Jingyuan Wang$^{1, \ast}$, Xin Lin$^2$, Yuxi Liu$^3$, Qilegeri$^2$, Kai Feng$^{4}$, Hui Lin$^{5}$}
\affiliation{%
  \institution{$1.$ Beijing Advanced Innovation Center for BDBC, Beihang University, Beijing, China $\ast$ Corresponding author.}
  \institution{$2.$ State Key Laboratory of Software Development Environment, Beihang University, Beijing, China}
  \institution{$3.$ College of Science and Engineering, Flinders University, Adelaide, Australia}
  \institution{$4.$ MOE Engineering Research Center of ACAT, School of Computer Science Engineering, Beihang University}
  \institution{$5.$ China Academy of Electronics and Information Technology, Beijing, China}
}

\renewcommand{\shortauthors}{J. Wang, et al.}

\begin{abstract}
Since December 2019, A novel coronavirus (2019-nCoV) has been breaking out in China, which can cause respiratory diseases and severe pneumonia. Mathematical and empirical models relying on the epidemic situation scale for forecasting disease outbreaks have received increasing attention. Given its successful application in the evaluation of infectious diseases scale, we propose a Susceptible-Undiagnosed-Infected-Removed (SUIR) model to offer the effective prediction, prevention, and control of infectious diseases. Our model is a modified susceptible-infected-recovered (SIR) model that injects undiagnosed state and offers pre-training effective reproduction number. Our SUIR model is more precise than the traditional SIR model. Moreover, we combine domain knowledge of the epidemic to estimate effective reproduction number, which addresses the initial susceptible population of the infectious disease model approach to the ground truth. These findings have implications for the forecasting of epidemic trends in COVID-19 as these could help the growth of estimating epidemic situation.
\end{abstract}





\maketitle

\section{Introduction}
In late December 2019, several local health institutions reported that the South China seafood wholesale market in Wuhan, one of the central cities of China, was epidemiologically related to the group of patients with unexplained pneumonia, and the global attention shifted to China \cite{who1}. Local health authorities identified a novel coronavirus, tentatively named 2019-nCoV, which is the third human cross-infection of coronavirus in 30 years, causing global health concerns \cite{whoglobal}. Chinese government took extraordinary measures, first in Wuhan and then in 12 other Chinese cities, to control the epidemic by closing markets and imposing blockades \cite{2}. The disease has now spread globally, including cases confirmed in 216 Countries \cite{3}. Depending on the World Health Organization statistics, as of 24 July, 2020, the Global region was 15,296,926 confirmed cases, and 628,903 cases died \cite{4}, surpassing the epidemic of the severe acute respiratory syndrome (SARS) in 2003 \cite{who2}. Global countries are issue policies to control the 2019-nCoV spread and provide financial support and health rescue. Therefore, predicting the future growth trend of the epidemic situation performs a vital function in measuring the large scale epidemic situation \cite{5}.

Mathematical and Empirical models have been widely adopt in the field of the epidemic, which adequately illustrates the transmission speed, spatial range, transmission path, and dynamic mechanism of infectious diseases \cite{6}. Depending on the categories of infectious diseases, conventional infectious disease models divide into SIR, time delay SIR, and SEIR model \cite{7,13,12}. Subsequently, there are many applications based on the epidemic model. Huo and Zhao considered birth and death rates on heterogeneous complex networks, which proposed a fractional SIR model and obtain results that when the disease-free equilibrium is globally asymptotically stable, the disease can disappear \cite{19}. A detailed study of the T-SIR (Time-series Susceptible-Infected-Recovered) model by Ottar et al. \cite{17} exposed the epidemic cycle and the outbreak of measles. A significant SIR model on the subject was presented by Jiang and Wei \cite{18}, which established the existence of Hopf bifurcations at the endemic equilibrium. Analysis of the SIR model involved with nonlinear incidence rate and time delay was proved that the underlying reproductive number $\mathcal{R}_0$ $\textgreater$ 1, the system is permanent \cite{21}. McCluskey further developed a SIR model of disease transmission with delay and nonlinear incidence on \cite{21}, which used a Lyapunov functional shown the global dynamics are entirely determined \cite{20}. A qualitative study by Li et al. \cite{22} identified A threshold $\sigma$ to determine the outcome of the disease on the SEIR model of infectious disease transmission. Smith et al. applied a latent period, excess death of the infected, and a standard incidence on the SEIR epidemic model to identify the reproduction number $\mathcal{R}_0$ \cite{23}.

However, these technologies have consistently shown a lack of investigation on disease characteristics, such as undiagnosed infectiousness. Therefore, it is critical to assess the effects of undiagnosed states on the epidemic progression for the benefit of global expectation. We are curious about the development of the epidemic and hope to contribute to its control. We investigated the global epidemic situation of COVID-19 infection and predicted future growth trends. Consequently, we proposed a mathematical model to analyze and forecast the number of individuals infected and recovered (including deaths) of COVID-19 in epidemic countries. This paper proposed a Susceptible-Undiagnosed-Infected-Removed (SUIR) Model and applies the simplex algorithm on the historical incidence data to fit $\beta$ and $\gamma$ of parameters, and predict the number of infected and removed (including cured and dead) in the next $t$ (per day). The contributions of this work are presented as follows: Previous studies of $\mathcal{R}(t)$ have not dealt with preprocessing of epidemic model. Most of these studies have suffered from the results of experiments in which the initial number of susceptible population is too large. To address this initialization challenge, the experimental work presented here provides the first pre-training approach ($\mathcal{R}(t)$) into how combined with the domain knowledge of epidemics, which enable applying pre-training $\mathcal{R}(t)$ to estimate the initial number of susceptible population. What's more, This is the first study to undertake an analysis of infectivity of undiagnosed individuals, which provided an important opportunity to advance the accuracy model's understanding. 

This paper is organized as follows. In Section 2, the SUIR epidemic model is formulated. In Section 3, the properties of database are studied, the basic Hyperparameters are given, the reliability of the solution and prediction error are illustrated. Section 4 highlights simulation results which conduct key control policies on the SUIR model. In Section 5, some statistical inferences and model results are discussed. Some conclusions are summarized in Section 6.

\section{Methodology}
\subsection{Dynamics model of infectious diseases}
There are a large number of published studies that describe the link between mathematical dynamic model and infectious diseases. Kermack-McKendrick \cite{7} proposed the system dynamic (Susceptible-Infective-Removal, SIR) model, which describes the transmission process of infectious diseases through a quantitative relationship to the transmission mechanism of general infectious diseases, analyzed the change rule of the number of infected cases and reveals the growth trend of infectious diseases. SIR model divided into three categories population, which include Susceptible ($S$), Infectious ($I$), and Removed ($R$). Based on the research of Kermack and McKendrick \cite{7}, Beretta and Takeuchi \cite{13} further developed their theories and proposed a time delay SIR model. Cooke and Driessche \cite{12} investigated the incubation period in the spread of infectious diseases, introduced "Exposed, $E$," and proposed a time delay model. Depending on the above infectious disease models, which were similar to the 2019-nCoV epidemic. This paper focuses on applying the SIR model as a fundamental hypothesis.

The traditional SIR model is based on the SI model \cite{14} to consider the recovery process of patients further and incorporates two critical parameters, including the infection rate ($\beta$) and the proportion coefficient ($\gamma$). Due to the characteristics of the 2019-nCoV, one of the main obstacles of the traditional SIR model, which not consider the infectivity of undiagnosed cases ($U$ state). Therefore, we inject the state of $U$ (undiagnosed) into the SIR model and propose the SUIR model to effectively track the spread of the epidemic situation and predict the future infection population.
\subsection{Differential Equations for Traditional SIR Model}
Investigations such as that conducted by \cite{14} have shown that the differential equation of SIR model and prerequisite (considering the recovery process of patients):
\begin{align}
\small
  \frac{\mathrm{d}S}{\mathrm{d}t} &= -\beta I S, \label{eq:1}\\
  \frac{\mathrm{d}I}{\mathrm{d}t} &= \beta I S - \gamma I, \label{eq:2}\\
  \frac{\mathrm{d}R}{\mathrm{d}t} &= \gamma I, \label{eq:3}\\
  S(t) + I(t) + R(t) &= M.\label{eq:4}
\end{align}
% $\displaystyle\frac{\mathrm{d}S}{\mathrm{d}t}$ = -$\beta$$I$$S$, (1)\\
% $\displaystyle\frac{\mathrm{d}I}{\mathrm{d}t}$ = $\beta$$I$$S$ - $\gamma$$I$, (2)\\
% $\displaystyle\frac{\mathrm{d}R}{\mathrm{d}t}$ = $\gamma$$I$, (3)\\
% $S(t)$ + $I(t)$ + $R(t)$ = $M$. (4)\\
$M$ represents the total population. Kermack-McKendrick \cite{7} assumed that patients obtain permanent immunity from recovery. Therefore, recovery patients can be removed from the system. In fatal infectious diseases, death cases were incorporated into the $R$ category.
\subsection{Structure and Hypothesis for SUIR model}
The SUIR model divides the population into four states: $S$, $U$, $I$ and $R$. Conceptually, the chain of the state transition is shown as Figure 1:

\begin{enumerate}
  \item $S$, susceptible;

  \item $U$, undiagnosed, refers to the patients who have been infected with diseases but have not been confirmed;

  \item $I$, (the confirmed cases not quarantined, infectious) and $IS$ (confirmed and quarantined, noninfectious), refers to the infected individual and was regarded as the confirmed patient;

  \item $R$, removed, refers to the patient who recovers with immunity or dies.

\end{enumerate}

\begin{figure}
  \centering
  \includegraphics[width=.7\columnwidth]{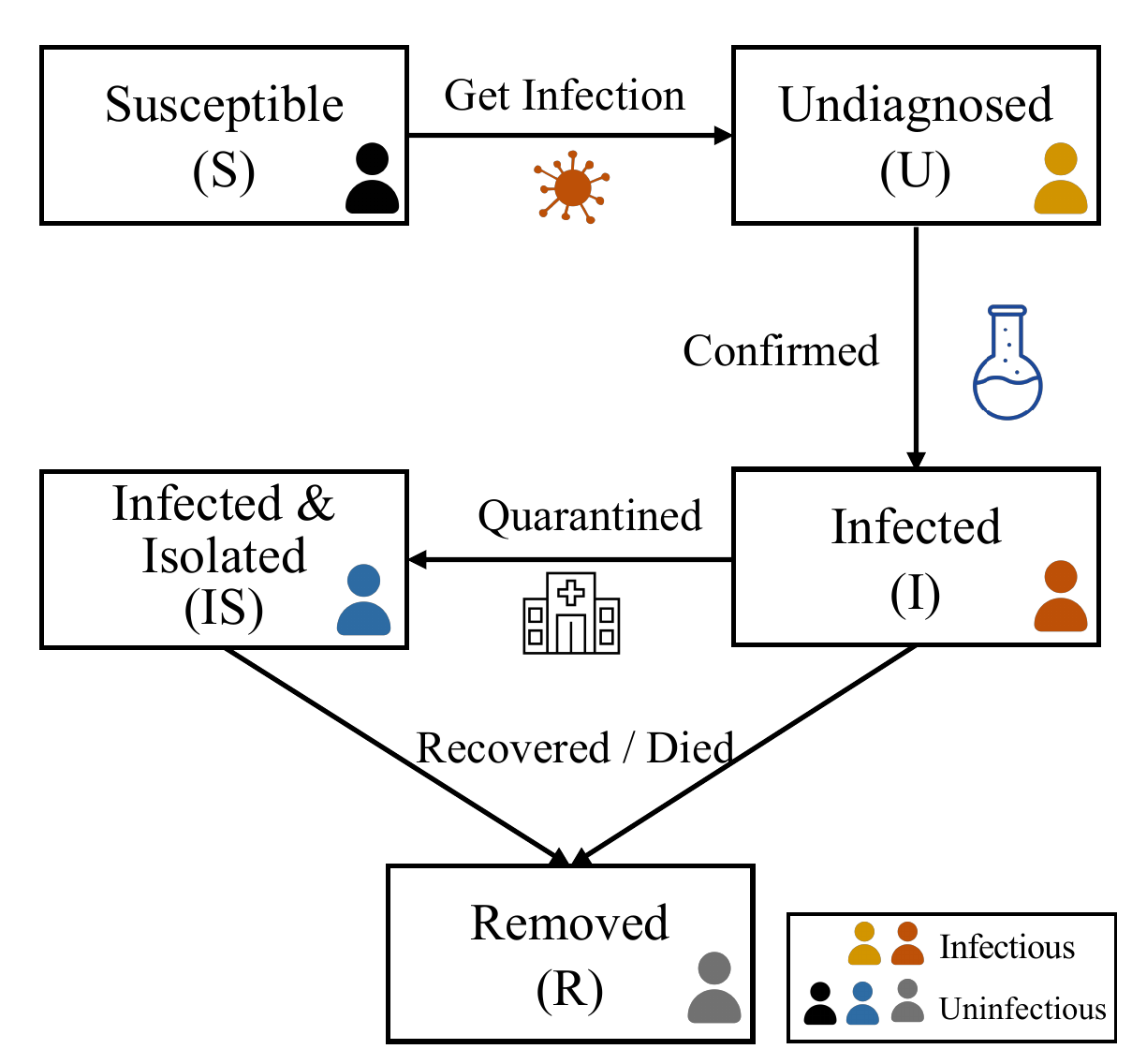}
  \caption{The chain of the state transition ($S$, $U$, $I$ , $IS$ and $R$)}
  \label{fig:chain}
\end{figure}
The first step in this process was to apply the historical data of $S$, $I$, and $R$ states to fit the transfer proportion parameters among population estimates. The second step used fitted parameters to predict the future disease state. The above model is similar to SIR, but the state of $I$ was divided into two sections in practical application, including the state of $I$ (Confirmed but not quarantined, infectious) and the state of $IS$ (Confirmed and quarantined, noninfectious).

Here, we assume that:

\begin{enumerate}
  \item The study area's total population never changes by time; both natural birth rates and mortality rates are not considered.

  \item The number of susceptible individuals affected by infectious diseases changes in direct proportion to the number of susceptible and infectious individuals.

  \item The growth rate of the number of quarantined and removal individuals is proportional to the number of infected individuals.

  \item Both the diagnosed (without quarantined) and the undiagnosed can infect individuals.

  \item The quarantined confirmed case can not infect individuals.

  \item Track close contact with confirmed cases and quarantine some undiagnosed individuals and assume that the average number of quarantined undiagnosed individuals caused by one confirmed case is $\rho$.
\end{enumerate}

\subsection{Define SUIR Model HyperParameters and Differential Equations}
In the modeling phase of the study, the initial susceptible population ($S_0$) was suggested to estimate. In this regard, we assumed that $S_0$ is the population base of the country or city to be estimated. However, there is a certain drawback associated with the use of the country or city's population base, which the entire population would consider as susceptible individuals. To address the challenge of $S_0$ initialization, Cintron et al. \cite{cintron2008estimation} offered the $\mathcal{R}(t)$ HyperParameter, which represents the average number of secondary cases of disease caused by a single infected individual over his or her infectious period, and note that $\mathcal{R}_0 = \mathcal{R}(0)$ for the initial day. For the estimation of $S_0$, a pretraining $\mathcal{R}(t)$ approach was conducted on the SIR model, in which the number of peak infections (the total number of peak $I$ and $R$) obtained from the experiment was regarded as initial $S_0$.

Here, we utilized two approaches to locate the pretraining $\mathcal{R}(t)$ sequence. We adopt China's data to explore the variation of $\mathcal{R}(t)$ with temperature and humidity and applied it to other countries' $\mathcal{R}(t)$ sequence estimation. Criteria for selecting the subjects were as follows: 100 Chinese cities with more than 40 confirmed cases were decided, and the $\mathcal{R}(t)$ series of these cities were estimated based on the historical incidence data using the \cite{wallinga2004different}, and the interval range was January 21 to January 23. Chinese government published that strict control measures would be implemented on January 23 \cite{tian2020investigation}; therefore, the $\mathcal{R}(t)$ sequence before January 23 reflects some extent of the epidemic situation (without intervention). To identify temperature and humidity patterns, the database applied the China Meteorological Data Service Center and chose the daily average temperature and relative humidity of the above 100 cities from January 21 to January 23. The fixed-effects (FE) panel regression \cite{bruderl2015fixed} was adopt according to the above procedure:
\begin{equation}
\label{eq:21}
y_{it} = coef_i + \mathbf{x}_{it} \mathbf{\beta} + v_{it},
\end{equation}
% $y_{it}$ = coef + $\mathbf{x}_{it}$$\mathbf{\beta}$ + $v_{it}$. (21)
where $y_{it}$ represents the effective reproduce number $\mathcal{R}$ of city $i$ in day $t$, $\mathbf{x}_{it}$ is the $(1 \times 2)$ vector, including temperature and relative humidity of city $i$ in day $t$. $\mathbf{\beta}$ represents a $(2 \times 1)$ vector of parameters, $coef_i$ represents the intercept of city $i$. $v_{it}$ is the error of city $i$ in day $t$.

To address the equation between $\mathcal{R}(t)$ and temperature and humidity, the following steps were taken: ordinary least squares regression has been used to investigate Eq.~\eqref{eq:21}, average of all $coef_i$ is taken as the final $coef$. The result is addressed as
\begin{equation}
\label{eq:22}
\mathcal{R}(t) = coef + \beta_1 \cdot Temperature(t) + \beta_2 \cdot RelativeHumidity(t).
\end{equation}
we defined the $(coef, \beta_1, \beta_2)$, which estimated from Chinese data. On completion of parameters, we adopt temperature and relative humidity data of countries to be estimated, in which the process of $\mathcal{R}(t)$ estimation was carried out Eq. \eqref{eq:22}.

The second method assumes that other countries adopt the same control measures as Wuhan, which used the historical incidence rate of Wuhan to estimate $\mathcal{R}(t)$ \cite{wallinga2004different}.

Due to the Eq. \eqref{eq:1}, the relationship among the $\mathcal{R}_0$, $\beta$ and $\gamma$ is
\begin{equation}
\small
\label{eq:5}
\mathcal{R}_0 = \frac{\beta \cdot M}{\gamma},
\end{equation}
based on $\mathcal{R}(t) = \frac{S(t)}{M}\mathcal{R}_0$ and an assumption that $S(t) \approx M$ in a short period. 
Consequently, the experiment was conducted under conditions in which $\mathcal{R}(t)$ and $\gamma$ were specified and lasted for $T$ days. Finally, $S_0$ was a total of $I(T)$ and $R(T)$.
\begin{equation}
\small
\label{eq:6}
I(T) + R(T) = S_0.
\end{equation}
% $I(T)$ + $R(T)$ = S$_0$. (6)

SUIR model includes six parameters:
\begin{itemize}

  \item $\beta$: The infection rate of contact between diagnosed and susceptible population.

  \item $\sigma$: The infection rate of undiagnosed cases in contact with the susceptible population.

  \item $\rho$: The average quarantine number of undiagnosed close contacts with confirmed cases.

  \item $\varepsilon$: The probability of undiagnosed infection by confirmed.

  \item $\lambda$: The probability of quarantine of confirmed cases.

  \item $\gamma$: The probability of removal of confirmed cases. (cure and death)
\end{itemize}
Differential Equations of SUIR model:
\begin{align}
\small
\frac{\mathrm{d}S}{\mathrm{d}t} =& -\beta S I - \sigma \cdot S \cdot \max((U - \rho \cdot IS)), 0), \label{eq:7} \\
\frac{\mathrm{d}U}{\mathrm{d}t} =& \beta S I + \sigma \cdot S \cdot \max((U - \rho \cdot IS)), 0) - \epsilon \cdot U, \label{eq:8}\\
\frac{\mathrm{d}I}{\mathrm{d}t} =& (1 - \lambda) \cdot \varepsilon \cdot U - \gamma \cdot I, \label{eq:9}\\
\frac{\mathrm{d}IS}{\mathrm{d}t} =& \lambda \cdot \varepsilon \cdot U - \gamma \cdot IS, \label{eq:10}\\
\frac{\mathrm{d}R}{\mathrm{d}t} =& \gamma \cdot (I + IS), \label{eq:11}
\end{align}
% $\displaystyle\frac{\mathrm{d}S}{\mathrm{d}t}$ = -$\beta$$S$$I$ - $\sigma$ * $S$ * $max$(($U$ - $\rho$ * $IS$)), 0), (7)\\
% $\displaystyle\frac{\mathrm{d}U}{\mathrm{d}t}$ = $\beta$$S$$I$ + $\sigma$ * $S$ * $max$(($U$ - $\rho$ * $IS$)), 0) - $\epsilon$ * $U$, (8)\\
% $\displaystyle\frac{\mathrm{d}I}{\mathrm{d}t}$ = ($1$ - $\lambda$) * $\varepsilon$ * $U$ - $\gamma$ * $I$, (9)\\
% $\displaystyle\frac{\mathrm{d}IS}{\mathrm{d}t}$ = $\lambda$ * $\varepsilon$ * $U$ - $\gamma$ * $IS$, (10)\\
% $\displaystyle\frac{\mathrm{d}R}{\mathrm{d}t}$ = $\gamma$ * ($I$ + $IS$), (11)\\
% The total number of population is $M$. $S(t)$ + $I(t)$ + $IS(t)$ + $R(t)$ = $M$. (12)
% The SUIR model is implemented by an overall flowchart (see Algorithm 1).
In this study, an investigation unit represents one day. The transition relationship between susceptible $(S)$ and undiagnosed $(U)$ for one day includes two categories. One is obtaining the infection from the infected individuals $(I)$, which represents the $\beta SI$ in Eq.~\eqref{eq:7}, similar to the traditional SIR model, the other is to obtain the infection from contact with undiagnosed individuals $(U)$, with a rate of $\sigma$. It should be noted that we assumed that the quarantine $\rho$ (undiagnosed individuals) could be quarantined by tracking close contact with the confirmed cases; therefore, the number of undiagnosed infections is $\sigma \cdot S \cdot max((U - \rho \cdot IS)), 0)$. In each investigation unit, $\varepsilon \cdot U$ (undiagnosed individuals) were diagnosed with a quarantine rate of $\lambda$. Therefore, the total newly diagnosed individuals ($(1 - \lambda) \cdot \varepsilon \cdot U$) was converted into Infected ($I$) individuals, and $\lambda \cdot \varepsilon \cdot U$ was converted into Infected \& quarantined ($IS$) individuals, as Eq.~\eqref{eq:9} and \eqref{eq:10}. The final stage of the study comprised a similar structure with the SIR model that the transition probability between confirmed and removal cases (cured or death) is $\gamma$, as Eq.~\eqref{eq:11}.

We defined $S(t)$, $U(t)$, $I(t)$, $IS(t)$, $R(t)$ as the number of susceptible, undiagnosed, infected, infected-isolated and removal individuals at time t, $\bigtriangleup t$ represent the unit time, the differential equations are summarized:
\begin{align}
\small
  S(t+\bigtriangleup t) =&~ S(t) - \beta S(t) I(t) \nonumber\\&-  \sigma S(t)\max((U(t) - \rho IS(t), 0) \label{eq:12}\\
  U(t+\bigtriangleup t) =&~  U(t) + \beta S(t) I(t) \nonumber\\&+  \sigma S(t)\max((U(t) - \rho IS(t), 0) - \varepsilon U(t) \label{eq:13}\\
  I(t+\bigtriangleup t) =&~  I(t) + (1 - \lambda) \varepsilon U(t) - \gamma I(t) \label{eq:14}\\
  IS(t+\bigtriangleup t) =&~ IS(t) + \lambda \varepsilon U(t) - \gamma IS(t) \label{eq:15}\\
  R(t+\bigtriangleup t) =&~ R(t) + \gamma(I(t)+IS(t)) \label{eq:16}
\end{align}

The total number of population is M,
\begin{equation}
% \small
\label{eq:17}
S(t) + U(t) + I(t) + IS(t) + R(t) = M,
\end{equation}
% $S(t)$ + $U(t)$ + $I(t)$ + $IS(t)$ + $R(t)$ = $M$, (17)
The cumulative confirmed cases
\begin{equation}
% \small
\label{eq:18}
C(t) = I(t) + IS(t) + R(t),
\end{equation}
% \small
The number of treated patients
\begin{equation}
% \small
\label{eq:19}
A(t) = I(t) + IS(t).
\end{equation}

The SUIR model is implemented by an overall flowchart (see Algorithm 1).

\begin{algorithm}[h]
\small
\caption{: SUIR Model}
\begin{algorithmic}[1]
\REQUIRE $\mathcal{R}(t)$, $(\beta_0, \sigma_0, \rho_0, \varepsilon_0, \lambda_0, \gamma_0)$, cumulative confirmed $\hat{C}(t)$, removal $\hat{R}(t)$, $T$
\ENSURE $S(t), U(t), I(t), IS(t), R(t)$
\STATE \textbf{Initialization: } $\beta_0, \sigma_0, \rho_0, \varepsilon_0, \lambda_0, \gamma_0$
\STATE \textbf{Pretraining $S_0$: } Apply $\mathcal{R}(t), \gamma$ on Eq.~\eqref{eq:1}, \eqref{eq:2}, \eqref{eq:3} and \eqref{eq:5}, \\
~~~obtain $S_0$ from Eq.~\eqref{eq:6}
\STATE \textbf{Estimation: } \\
\STATE ~~Apply $(\beta_0, \sigma_0, \rho_0, \varepsilon_0, \lambda_0, \gamma_0)$ on Eq.~\eqref{eq:12}-\eqref{eq:16}, obtain $C(t)$ and $R(t)$ from Eq.~\eqref{eq:16} and \eqref{eq:18}
\STATE ~~~Obtain MSE of $C(t)$ and $\hat{C}(t)$, $R(t)$ and $\hat{R}(t)$
\STATE ~~~Solve $(\beta, \sigma, \rho, \varepsilon, \lambda, \gamma)$ by using Nelder-Mead solver to minimize MSE
\STATE \textbf{Simulation: }
\FOR {$t = 1$ to $T$}
\STATE Apply $(\beta, \sigma, \rho, \varepsilon, \lambda, \gamma)$ on Eq.~\eqref{eq:12}-\eqref{eq:16}, \\update $S(t), U(t), I(t), IS(t)$ and $R(t)$
\ENDFOR
\end{algorithmic}
\end{algorithm}

This deterministic epidemic model is based on the hypothesis of 1, 2, 3, 4, 5, and 6; therefore, it will lose accuracy in other infectious diseases. But in 2019-nCoV with a high probability of the correct hypothesis.
\section{Results}
\subsection{Database description}
Data were gathered from multiple sources at various time points during the epidemic outbreak. To investigate the prediction of China, we adopted the National Health Commission (NHC) of China daily epidemic statistics report \cite{24}, which composed of the cumulative number of infectious, recovered, and death cases in China and summarized in Table 1 (e. g. Wuhan). To address the global outbreak, the JHU CSSE Database \cite{25} are summarized in Table 2 (e. g. USA). Since January 22, 2020, the JHU CSSE Database was updated per day and composed of three sections, including the cumulative number of infectious, recovered, and death cases. Furthermore, we select 15-days data at least before the forecast date for training the model, \textit{e.g.}, to obtain trend of the USA on April 1, we collects the data of USA from March 17 to March 31.
\begin{table}[htbp]
  \centering
  \small
  \caption{Historical Data of Wuhan Released by the National Health Commission of China}
    \begin{tabular}{cccc}
    \toprule
    Date & \multicolumn{1}{p{5.415em}}{Cumulative } & \multicolumn{1}{p{5.415em}}{Cumulative } & \multicolumn{1}{p{5.415em}}{Cumulative } \\
         & Infectious & Recovered & Deaths \\
    \midrule
    1/27/20 & 1590 & 47   & 85 \\
    1/28/20 & 1905 & 47   & 104 \\
    1/29/20 & 2261 & 51   & 129 \\
    1/30/20 & 2639 & 72   & 159 \\
    1/31/20 & 3215 & 123  & 192 \\
    2/1/20 & 4109 & 155  & 224 \\
    2/2/20 & 5142 & 166  & 265 \\
    2/3/20 & 6384 & 303  & 303 \\
    2/4/20 & 7828 & 368  & 362 \\
    2/5/20 & 10117 & 431  & 414 \\
    2/6/20 & 11618 & 534  & 478 \\
    2/7/20 & 13603 & 698  & 545 \\
    2/8/20 & 14982 & 877  & 608 \\
    2/9/20 & 16902 & 1044 & 681 \\
    2/10/20 & 18454 & 1206 & 748 \\
    \bottomrule
    \end{tabular}%
  \label{tab:nhc-wuhan}%
\end{table}%
% Table generated by Excel2LaTeX from sheet 'Sheet2 (2)'
\begin{table}[htbp]
  \centering
  \small
  \caption{Historical Data of USA Released by JHU CSSE Database}
    \begin{tabular}{cccc}
    \toprule
    Date & \multicolumn{1}{p{5.5em}}{Cumulative } & \multicolumn{1}{p{5.5em}}{Cumulative } & \multicolumn{1}{p{5.5em}}{Cumulative } \\
         & Infectious & Recovered & Deaths \\
    \midrule
    3/17/20 & 6420 & 74   & 105 \\
    3/18/20 & 7769 & 74   & 118 \\
    3/19/20 & 13680 & 106  & 200 \\
    3/20/20 & 16638 & 147  & 233 \\
    3/21/20 & 24148 & 171  & 285 \\
    3/22/20 & 33276 & 178  & 417 \\
    3/23/20 & 43901 & 178  & 557 \\
    3/24/20 & 53740 & 348  & 706 \\
    3/25/20 & 66132 & 361  & 947 \\
    3/26/20 & 85486 & 713  & 1288 \\
    3/27/20 & 103942 & 870  & 1689 \\
    3/28/20 & 122666 & 1073 & 2147 \\
    3/29/20 & 142328 & 4767 & 2489 \\
    3/30/20 & 163429 & 5764 & 3008 \\
    3/31/20 & 188172 & 7024 & 3873 \\
    \bottomrule
    \end{tabular}%
  \label{tab:jhu-us}%
\end{table}%
\subsection{Hyperparameter simulations for the SUIR model}
\begin{figure}
  \centering
  \includegraphics[width=.8\columnwidth]{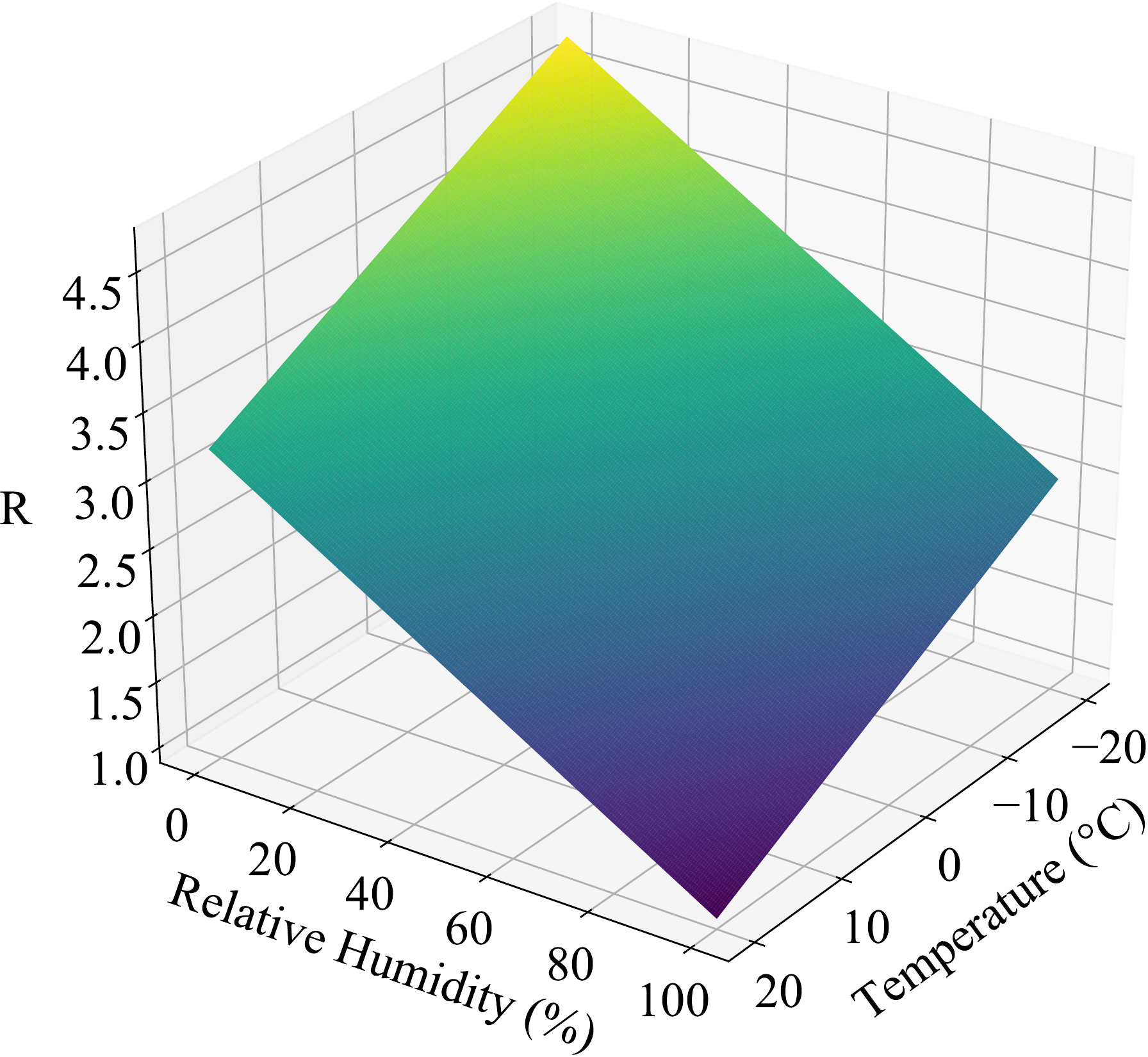}
  \caption{$\mathcal{R}(t)$ v.s. temperature and relative humidity}
  \label{fig:r0regression}
\end{figure}

\begin{figure}
  \centering
  \includegraphics[width=.78\columnwidth]{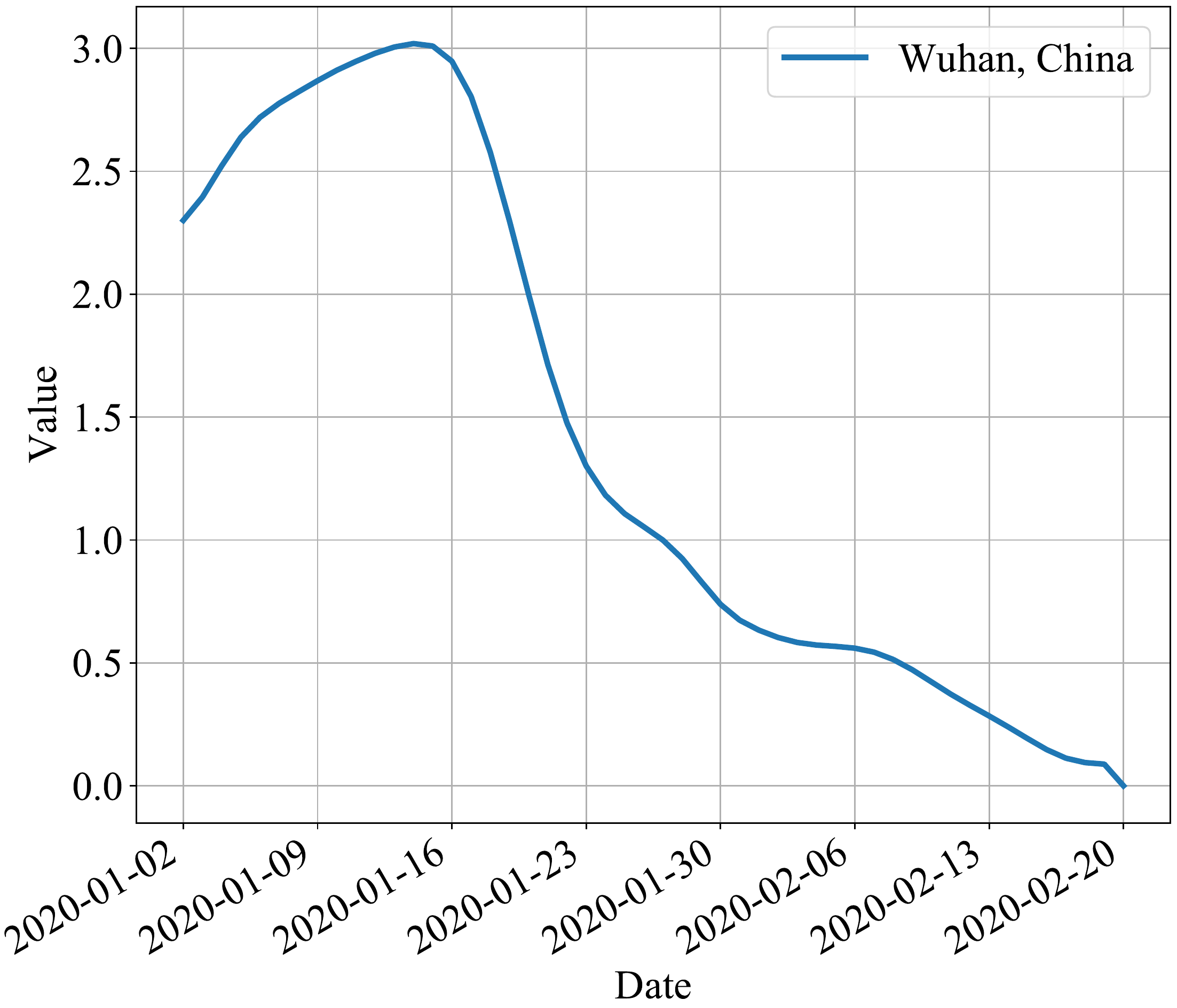}
  \caption{The infection dimension of confirmed cases under intervention ($\mathcal{R}(t)$) in Wuhan}
  \label{fig:r0wuhan}
\end{figure}

\begin{figure}
  \centering
  \includegraphics[width=.78\columnwidth]{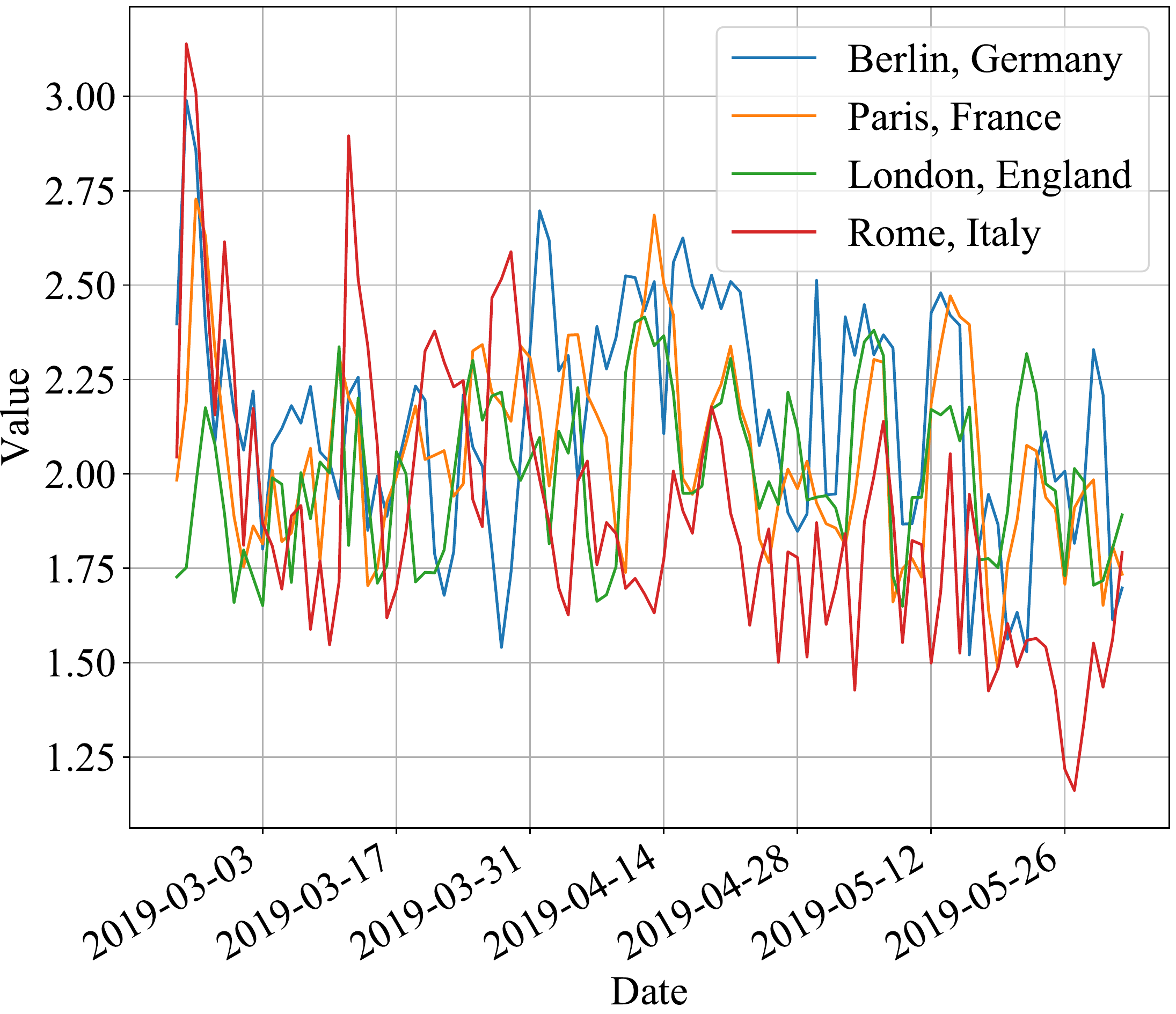}
  \caption{The infection dimension of confirmed cases ($\mathcal{R}(t)$) in outbreak countries}
  \label{fig:r0country1}
\end{figure}

The hyperparameter experiment's purpose was to adopted two sections of $\mathcal{R}(t)$ based on the SIR model, aiming to obtain the upper bound $S_0$. Simple statistical analysis was used to investigate Eq.~\eqref{eq:22}, which demonstrated the relationship between $\mathcal{R}(t)$ and temperature and humidity. Following statistical analysis, $coef$, $\beta_1$, and $\beta_2$ were obtained, which their values are 3.968, -0.0383 and -0.0224. Figure. \ref{fig:r0regression} presents the range of temperature and humidity for $\mathcal{R}(t)$. Figure. \ref{fig:r0wuhan} shows an overview of $\mathcal{R}(t)$ of Wuhan calculated by \cite{wallinga2004different}. Since January 28th, the $\mathcal{R}(t)$ estimated by the incidence rate of Wuhan has a significant decreasing trend and is lower than 1, which could be the consequence of Wuhan's control measures. Figure. \ref{fig:r0country1} provides the range of $\mathcal{R}(t)$ sequences for some outbreak countries. A clear trend of $\mathcal{R}(t)$ sequences under without intervention has widely fluctuated in this analysis. Table \ref{tab:twor0} provides the estimation results of $S_0$ obtained from two groups of $\mathcal{R}(t)$ functioning in different countries. The estimation experiments start on February 21st and last for 100 days.

% Table generated by Excel2LaTeX from sheet 'Sheet8'
% \begin{table}[htbp]
%   \centering
%   \small
%   \caption{$S_0$ of some countries obtained by two sections of $\mathcal{R}_0$}
%     \begin{tabular}{ccc}
%     \toprule
%     $S_0$ & From Wuhan's $\mathcal{R}_0$ & From Local $\mathcal{R}_0$ \\
%     \midrule
%     Italy & 51,100  & 161,000  \\
%     USA   & 78,000  & 345,000  \\
%     Iran & 40,000  & 84,000  \\
%     UK   & 29,000  & 86,000  \\
%     Spain & 71,000  & 180,000  \\
%     France & 26,000  & 147,000  \\
%     Germany & 35,000  & 130,000  \\
%     \bottomrule
%     \end{tabular}%
%   \label{tab:twor0}%
% \end{table}%

% Table generated by Excel2LaTeX from sheet 'Sheet8'
\begin{table}[htbp]
  \centering
  \small
  \caption{$S_0$ of some countries obtained by two sections of $\mathcal{R}(t)$}
    \begin{tabular}{ccc}
    \toprule
    $S_0$ & From Wuhan's $\mathcal{R}(t)$ & From Local $\mathcal{R}(t)$ \\
    \midrule
    Italy & 51,000  & 275,000  \\
    USA   & 78,000  & 6,628,000  \\
    Iran & 40,000  & 396,000  \\
    UK   & 29,000  & 319,000  \\
    Spain & 71,000  & 496,000  \\
    France & 26,000  & 357,000  \\
    Germany & 35,000  & 263,000  \\
    \bottomrule
    \end{tabular}%
  \label{tab:twor0}%
\end{table}%

\subsection{Solution of the SUIR model} 
\begin{figure}
  \centering
  \includegraphics[width=.83\columnwidth]{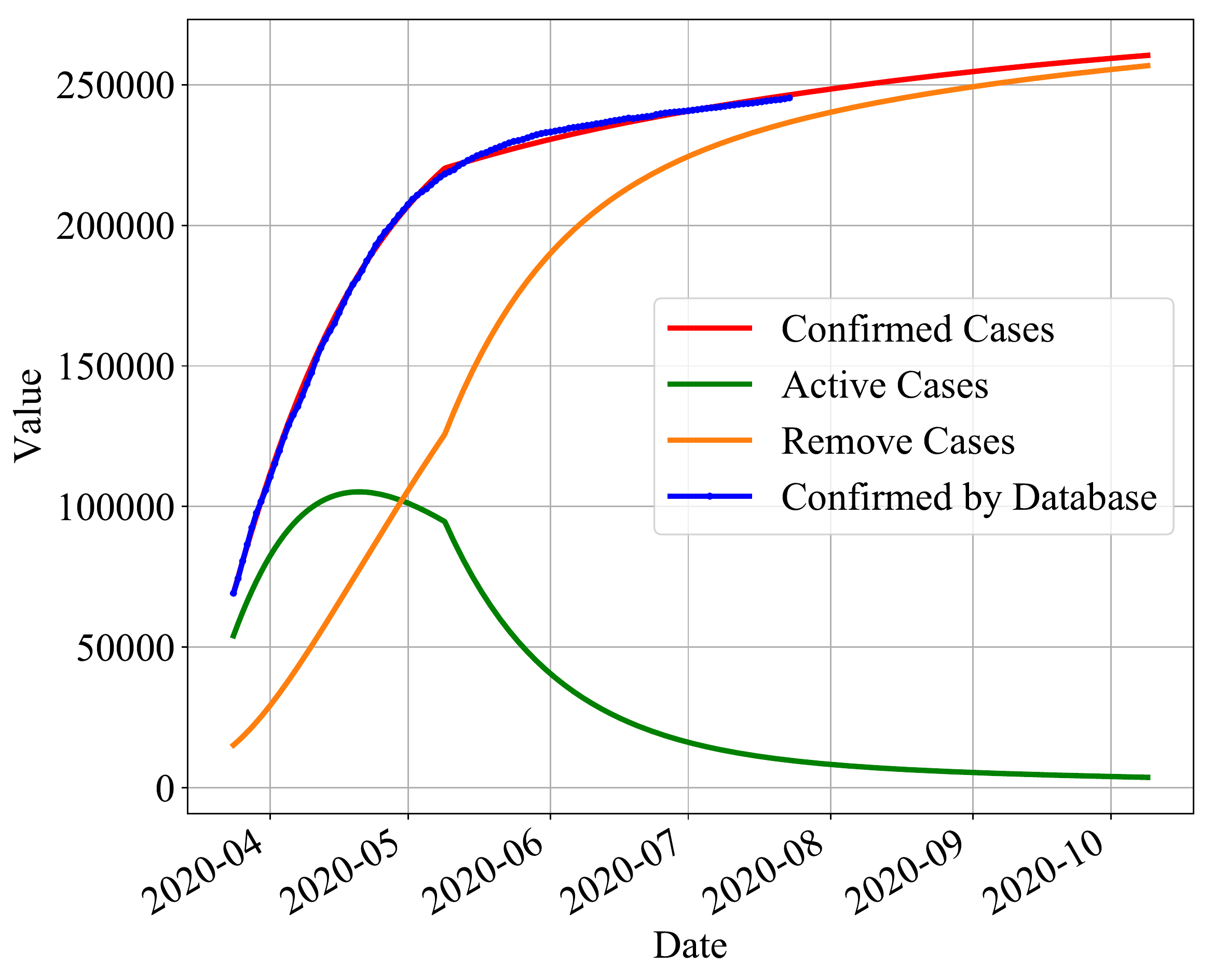}
  \caption{Prediction Result of Italy}
  \label{fig:Italy}
\end{figure}

\begin{figure}
  \centering
  \includegraphics[width=.83\columnwidth]{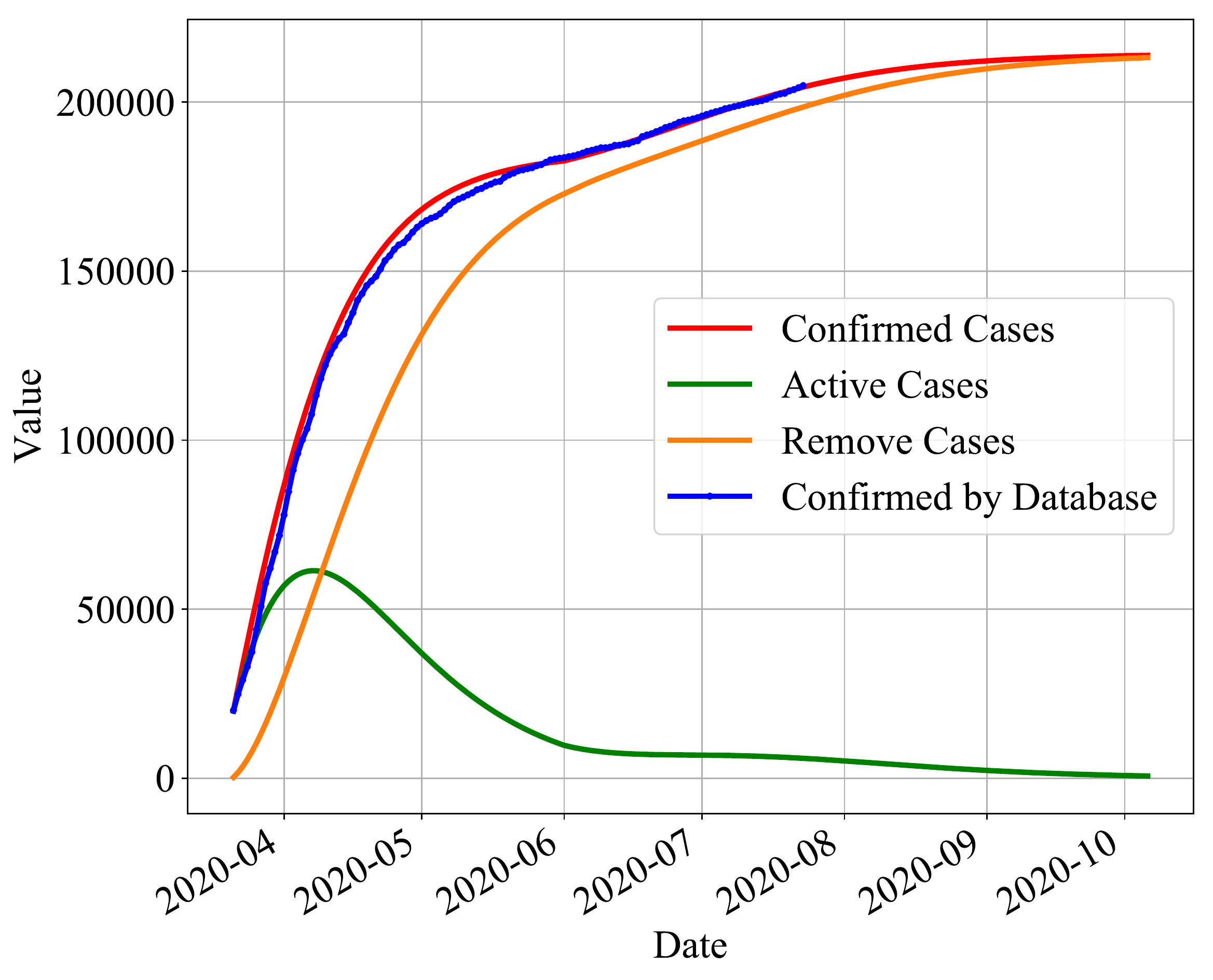}
  \caption{Prediction Result of Germany}
  \label{fig:Germany}
\end{figure}

To obtain fitted parameters, we collected the Historical data as of July 24, and Hyperparameter $S_0$ was estimated from local $\mathcal{R}(t)$ on the SUIR model. The results of fitted parameters are shown in Table \ref{tab:param}. Data from this table can be applied some parameters ($\beta$, $\sigma$, $\rho$, $\varepsilon$, $\lambda$, $\gamma$) on Eq.~\eqref{eq:12}-\eqref{eq:16} , where obtain the ($S$, $U$, $I$, $IS$, $R$) of outbreak countries in the future and estimate the cumulative number of confirmed cases and removed cases. Figure. \ref{fig:Italy} and Figure. \ref{fig:Germany} illustrate the summary statistics for Italy and Germany. As can be seen from Figure \ref{fig:Italy}, we can see that the number of confirmed cases in Italy showed an increasing trend before September, and then tended to be flat. The number of active cases first increased and then decreased with time, approaching a peak in mid-April, with approximately 100,000 cases. As shown in Figure \ref{fig:Germany}, further analysis showed that the number of confirmed cases in Germany revealed a growing trend before the middle of September, and active cases peaked were reported in April, including approximately 60,000 cases.

\begin{table}[htbp]
  \centering
  \small
  \caption{Fitted parameters of outbreak countries}
    \begin{tabular}{ccccccc}
    \toprule
    Country & $\beta$ & $\sigma$ & $\rho$ & $\epsilon$ & $\lambda$ & $\gamma$ \\
    \midrule
    Italy & 9.77E-06 & 7.54E-07 & 0.89  & 0.03  & 0.99 & 2.54E-02 \\
    USA   & 6.67E-06 & 1.38E-06 & 1.32  & 0.75  & 0.84 & 1.68E-03 \\
    Iran & 7.79E-05 & 1.17E-06 & 16.25  & 0.06  & 0.92 & 6.95E-02 \\
    UK   & 5.71E-06 & 1.26E-08 & 17.50  & 0.79  & 0.19 & 2.27E-02 \\
    Spain & 3.32E-06 & 1.92E-06 & 10.10  & 0.30  & 0.43 & 4.96E-02 \\
    France & 4.65E-05 & 1.37E-06 & 10.00  & 0.29  & 0.95 & 3.56E-02 \\
    Germany & 1.49E-03 & 9.37E-05 & 5.00  & 0.13  & 0.99 & 6.36E-02 \\
    \bottomrule
    \end{tabular}%
  \label{tab:param}%
\end{table}%

\subsection{Prediction Error of the SUIR model}
In this study, we assume that prediction date between $t$ and $T$, the error of prediction equation is defined
\begin{equation}
\label{eq:20}
\small
e(t, T) = \frac{|C(t+T) - \hat{C}(t+T)|}{\hat{C}(t+T)},
\end{equation}
where $C(t)$ denotes the cumulative confirmed cases predicted by SUIR model, and $\hat{C}(t)$ denotes the cumulative confirmed cases of the Database. Table \ref{tab:SUIRerror} compares the summary statistics for the prediction results of the SIR and SUIR models in outbreak countries. It is apparent from this table that the SUIR model's precision is much higher than that of the traditional SIR model.
Table \ref{tab:SUIRdroprate} summarizes the decrement rate of prediction error of SUIR model compared with the SIR model. Under all countries that investigated, the SUIR model achieves a 38.4\% lower prediction error than SIR model on average.

\begin{table*}[htbp]
  \centering
  \small
  \caption{Prediction error of SIR and SUIR model after $T$ days}
    \begin{tabular*}{\textwidth}{c @{\extracolsep{\fill}} cccccccc}
    \toprule
    Country & Model & \multicolumn{7}{c}{$T$} \\
         &      & 1    & 2    & 3    & 4    & 5    & 6    & 7 \\
    \midrule
    \multirow{2}[1]{*}{Italy} & SIR  & 0.94\% & 2.07\% & 3.13\% & 4.12\% & 4.88\% & 5.15\% & 5.02\% \\
         & SUIR & 0.43\% & 1.01\% & 1.49\% & 1.90\% & 2.25\% & 2.55\% & 2.73\% \\
    \multirow{2}[0]{*}{US} & SIR  & 2.07\% & 2.88\% & 3.06\% & 3.98\% & 4.80\% & 5.43\% & 6.83\% \\
         & SUIR & 2.02\% & 2.64\% & 2.69\% & 2.59\% & 2.78\% & 5.06\% & 6.62\% \\
    \multirow{2}[0]{*}{Iran} & SIR  & 5.00\% & 9.61\% & 13.56\% & 16.88\% & 19.72\% & 22.02\% & 23.90\% \\
         & SUIR & 1.61\% & 3.09\% & 4.64\% & 6.08\% & 7.31\% & 8.20\% & 8.83\% \\
    \multirow{2}[0]{*}{UK} & SIR  & 3.28\% & 5.66\% & 6.12\% & 6.31\% & 6.90\% & 7.06\% & 5.95\% \\
         & SUIR & 2.96\% & 5.36\% & 5.09\% & 3.50\% & 2.86\% & 2.35\% & 2.63\% \\
    \multirow{2}[0]{*}{Spain} & SIR  & 2.91\% & 5.74\% & 8.03\% & 9.53\% & 10.11\% & 10.07\% & 9.73\% \\
         & SUIR & 1.72\% & 2.71\% & 3.20\% & 3.85\% & 4.10\% & 4.13\% & 4.15\% \\
    \multirow{2}[0]{*}{France} & SIR  & 2.26\% & 4.38\% & 4.38\% & 4.74\% & 5.34\% & 9.26\% & 11.34\% \\
         & SUIR & 1.50\% & 2.47\% & 2.57\% & 4.18\% & 5.29\% & 8.43\% & 9.55\% \\
    \multirow{2}[1]{*}{Germany} & SIR  & 2.56\% & 4.51\% & 5.62\% & 6.13\% & 6.34\% & 5.90\% & 5.08\% \\
         & SUIR & 1.88\% & 3.35\% & 4.00\% & 4.66\% & 5.05\% & 4.93\% & 4.50\% \\
    \bottomrule
    \end{tabular*}
  \label{tab:SUIRerror}%
\end{table*}

% Table generated by Excel2LaTeX from sheet 'Sheet6'
\begin{table}[htbp]
  \centering
  \small
  \caption{7-day average decrement rate on prediction error of SUIR compared with SIR model}
    \begin{tabular}{cc}
    \toprule
    Country & Decrement Rate of Prediction Error \\
    \midrule
    Italy & 51.74\% \\
    USA   & 15.63\% \\
    Iran & 64.88\% \\
    UK   & 36.78\% \\
    Spain & 55.60\% \\
    France & 22.34\% \\
    Germany & 21.92\% \\
    \bottomrule
    \end{tabular}
  \label{tab:SUIRdroprate}%
\end{table}

\section{Simulation}

Multiple parameters of models reveal that the spreading status of epidemics. In an attempt to make adjustment on some parameters, which enables simulation implementing on the different intensities of control policies. This section presents two prototypes of simulations based on Wuhan's policies.

\subsection{Effect of close contact isolation}
For the purpose of controlling the epidemic situation, Wuhan has taken strict quarantine measures for close contacts, tracking the close contacts of confirmed cases, and conducted medical isolation observation, which process describes the change of the parameter $\rho$. For the simulation of the tracking process, the intensity of isolation measures can be adjusted to simulate the spread of the epidemic scale. The first step in this process was to adapt Wuhan data to fit the model, where the specific model parameters were obtained. The second step was to retain other parameters fixed and adjusted the isolation ratio of close contacts as $0.8\rho$, $0.6\rho$, $0.55\rho$, which enabled the estimation of the spread of the epidemic situation.

\begin{figure}
  \centering
  \includegraphics[width=.83\columnwidth]{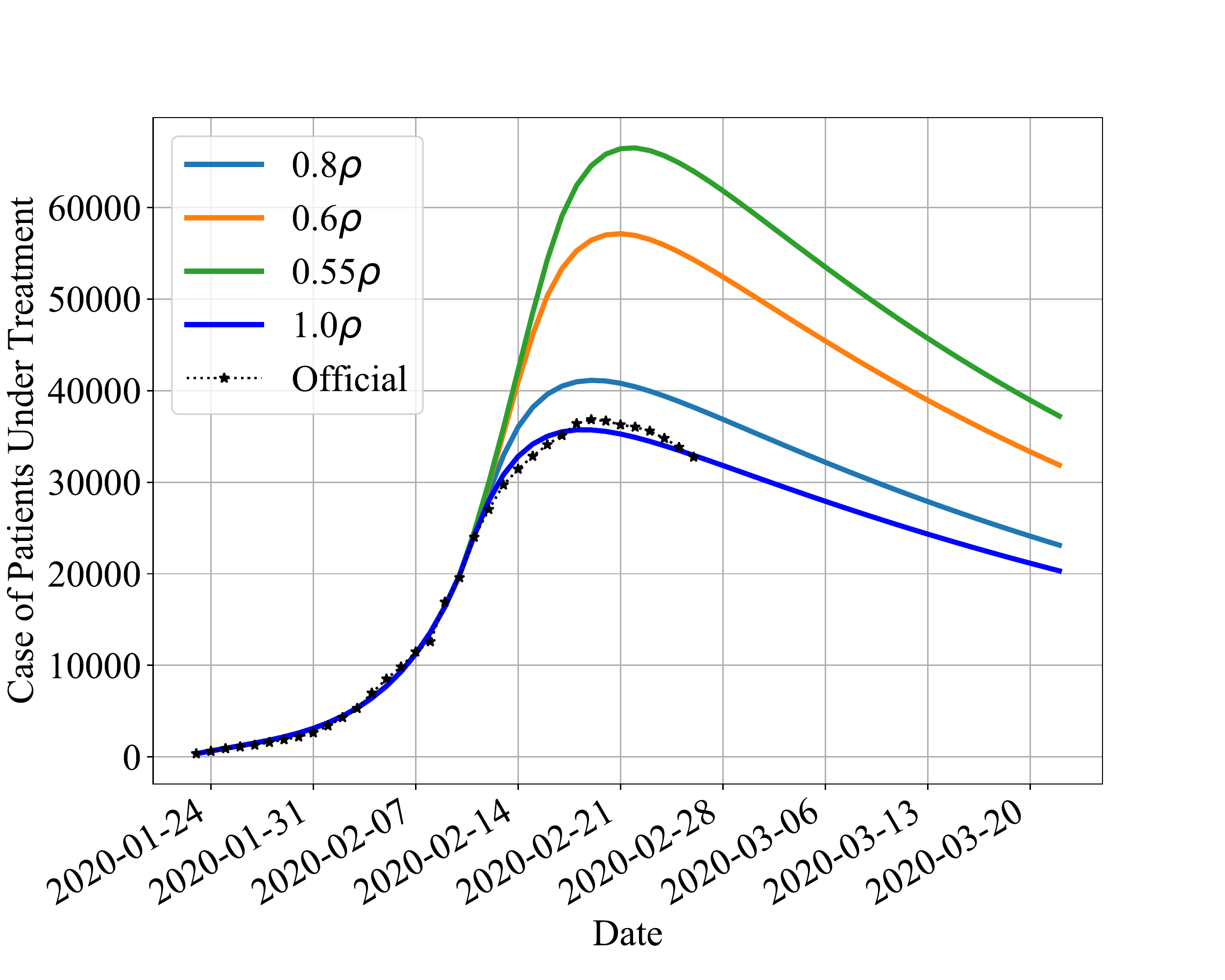}
  \caption{Under different isolation ratio, the number of patients in treatment changes with time}
  \label{fig:rho_active}
\end{figure}

Looking at Figure \ref{fig:rho_active}, it is apparent that when the isolation ratio decreases, the patients' peak number and time in treatment will change. The most interesting aspect of this graph is that when the isolation ratio decreased to $0.55\rho$, the peak number of patients in treatment reached more than twice the real value (Green curve). A clear benefit of tracking and isolating close contacts of the confirmed cases in the prevention of the epidemic's growth could be identified in this analysis.

\subsection{Effect of diagnosis rate}
On February 5, 2020, the government of Hubei Province issued policies to accelerate the diagnosis and measure the number of infected patients \cite{nanfangzhoumo1}. Delayed diagnose can cause infection patients to enable not be quarantined in time and generate more infected cases. In the SUIR model, parameter $\varepsilon$ represents the rate of diagnosis of the undiagnosed infected. The experiments were simulated using the Wuhan data to fit and obtained model parameters. After training, the parameters were remained unchanged except the diagnosis rate, which decreased to $0.8\varepsilon$, $0.6\varepsilon$, and $0.55\varepsilon$, and applied the estimation of epidemic spread scale.

\begin{figure}
  \centering
  \includegraphics[width=.83\columnwidth]{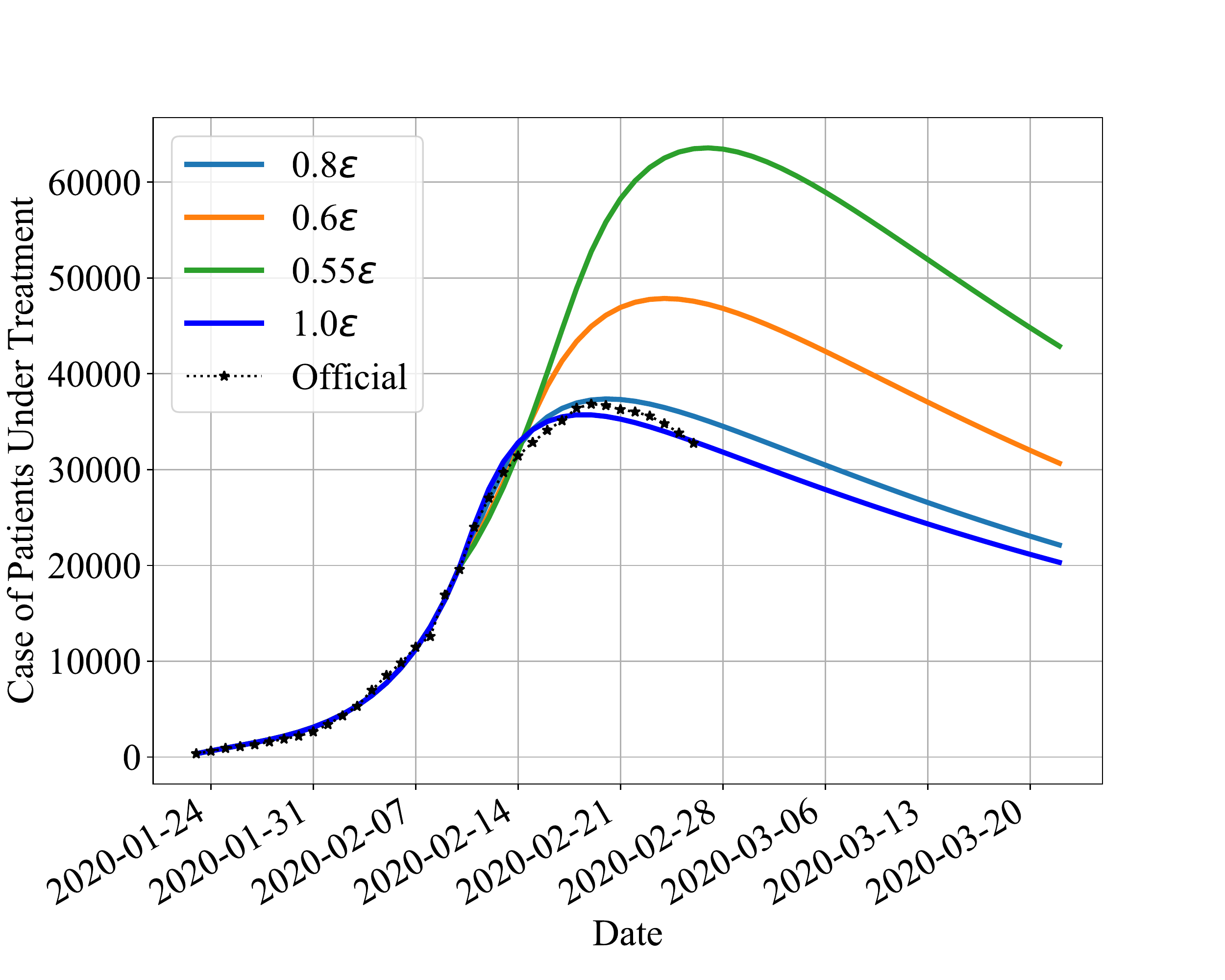}
  \caption{Under different diagnostic rates, the number of patients in treatment changes with time}
  \label{fig:epsilon_active}
\end{figure}

From the data in Figure \ref{fig:epsilon_active}, it is apparent that the rate of diagnosis decreased, which causes the curve's peak quantity and peak arrival time were variable. The delayed diagnosis rate will cause more infections, performing more significant pressure on medical resources.

\section{Discussion}
Our results suggest that the SUIR model's prediction accuracy is precise than the traditional SIR model in 2019-nCoV. On the question of initial dimension $S_0$ (susceptible individual) settings, this study discovered that applied $\mathcal{R}(t)$ sequence (includes domain knowledge) on the SIR model to pretraining, which contributes a reference for establishing $S_0$. These results in Table 3 highlight $S_0$ estimation results obtained by performing two sections of $\mathcal{R}(t)$ to different countries. It can be seen from the results in Table 3 that the $S_0$ from Wuhan $\mathcal{R}(t)$ sequence is smaller than that from the local $\mathcal{R}(t)$ sequence. A possible explanation for these results may be the lack of the same control measures as Wuhan, China. Consequently, national control measures will affect the infected individual trend.

The actual epidemic trend since our analyses has present the Hyperparameters has a significant weight on predict. More recent examples of narrative studies within the influence of air temperature and humidity on hyperparameter $\mathcal{R}(t)$ (COVID-19) can be found in the work of \cite{sajadi2020temperature} and \cite{15}. It is encouraging to compare Figure \ref{fig:r0regression} with that found by \cite{sajadi2020temperature} who found that low temperature and low humidity will make the $\mathcal{R}(t)$ of COVID-19 more significant, reflecting the virus's more solid transmission ability. A detailed study by Kermack-McKendrick \cite{7} indicates that the relationship among $\beta$, $\gamma$ and $\sigma$ is $\sigma$ = $\beta$ / $\gamma$. In this regard, $\sigma$ determines the spread of infectious diseases, which $\mathcal{R}_0$ $\textless$ 1 / $\sigma$ represents the transmission is limited \cite{27}. Our study generally supports \cite{7} and \cite{27} speculations; we believe the fluctuation of $\beta$ indicates the epidemic infectivity. Our results in Table \ref{tab:param} compares an overview of fitted parameters in seven countries, Iran, France and Germany have higher $\beta$ than other countries, indicating that 2019-nCoV more infectious in these countries. Furthermore, the most obvious finding to emerge from the analysis is that $\gamma$ of the United States located at a lower level than other countries, indicating that confirmed cases in the United States have a postponed treatment period than in other countries. Another important finding was that Italy and Iran have a low level of $\varepsilon$, which suggests there is a higher risk of undiagnosed infections at the early stage. As a result, hyperparameters determine the fitting quality and prediction precision.

Error statistics of the model validates the accuracy of the SUIR model. In this study, SIR model and SUIR model were used to 7-day prediction in multiple outbreaks countries (Table \ref{tab:SUIRerror}); the most obvious finding to emerge from the analysis is that the SUIR model can adapt to the characteristic of COVID-19 and minimize the estimation errors. These results reflect those of Cao et al. \cite{seirmodifed_cao} who also found that undiagnosed cases took transmission ability. Further analysis showed that the SUIR model and SIR model errors were increasing with the increment of prediction. The weak performance of the SUIR model is interesting, but not surprising. These possible sources of error could come from hyperparameter drive and a prolonged latent period.

Some limitations of our model are the initialization of the parameters, such as $\beta$ and $\gamma$. Under the improper setting, the model will not be able to fit the historical data well and cause a high prediction error. Another limitation of our study is that we did not account for the impact of imported cases. However, these problems could be solved if we apply empirical models to trial and error and often to update the data source. A further study with more focus on parameter optimization and data extraction is therefore suggested.

\section{Conclusion and Future Work}
In this paper, we propose a SUIR model and combine the characteristics of the 2019-nCoV epidemic to simulate and predict the future trend of the epidemic. The findings of this study suggest that our prediction model could precisely predict the number of infected individuals in 2019-nCoV under the database of \cite{25} and \cite{24} with a low error rate. The evidence from this study suggests that utilizing $\mathcal{R}(t)$ (domain knowledge) to predict the future trend of susceptible individuals has a significant effect observed from the model. The second major finding was that introducing the infectious state of undiagnosed cases into the model resulted in much higher prediction accuracy than the traditional SIR model. By adjusting some parameters of SUIR model, we are able to simulate the transmission of COVID-19 under different intensities of control policies, and evaluate the effects of them. These findings of this research provide insights for control epidemic situations. The most important limitation lies in the fact that until we complete this study, the 2019-nCoV epidemic in some continents, such as the Africa, has not recorded the outbreak stage. A further study could assess the long-term effects of recovery cases on infectiousness.
\begin{acks}
The work was supported by the National Key Research \& Development Program of China (Grant No. 2019YFB2102100), the National Natural Science Foundation of China (Grant No. 71531001, 61872369), the Fundamental Research Funds for the Central Universities (Grant No. YWF-20-BJ-J-839) and CCF-DiDi Gaia Collaborative Research Funds for Young Scholars.
\end{acks}

\bibliographystyle{ACM-Reference-Format}
\bibliography{cite}

\appendix









\end{document}